# A New Path to Nanoscale Cellular Analysis with Monochromated Electron Energy-Loss Spectroscopy

*Jordan A. Hachtel[1*], Jacek Jakowski[2], Jingsong Huang[1], Santa Jansone-Popova[3], Ilja Popovs[3], Elizabeth A. Richardson[4], Barbara R. Evans[3], Peter Rez[5], Eric V. Formo[4*]*

[1] Center for Nanophase Materials Sciences, Oak Ridge National Laboratory, Oak Ridge, TN, USA
[2] Computational Sciences and Engineering Division, Oak Ridge National Laboratory, Oak Ridge, TN, USA
[3] Chemical Sciences Division, Oak Ridge National Laboratory, Oak Ridge, TN, USA
[4] Georgia Electron Microscopy, University of Georgia, Athens, GA, USA
[5] Department of Physics, Arizona State University, Tempe, AZ, USA
[*] Correspondence to: hachtelja@ornl.gov
[*] Correspondence to: eformo@uga.edu

## Abstract

High-spatial-resolution vibrational spectroscopy is one of the principal techniques for nanoscale compositional analysis in biological materials. Here, we present a new method for the analysis of whole-cell biological specimens through nanoscale vibrational electron energy-loss spectroscopy (EELS) in the monochromated scanning transmission electron microscope. Using the combined spatial and spectral resolution of the technique, we examine the vascular system of a cucumber stem and identify clear physical and vibrational signatures from the different cellular regions with high spatial resolution. Furthermore, using first-principles calculations combined with optical and EELS spectroscopy on the individual components that make up the cucumber stem, we unravel the physical mechanisms of the vibrational signatures and directly assign compositional origins to the cell walls and bodies of different cellular regions. These results demonstrate that monochromated electron energy-loss spectroscopy is a promising technique for nanoscale spatial mapping of the chemical composition of biological materials.

## Main Text

The electron microscope (EM) has been an essential tool in the imaging and analysis of biological materials[1–3]. While many of these studies have focused on imaging, spectroscopic analysis through electron energy-loss spectroscopy (EELS) and energy dispersive spectroscopy (EDS) has extended the capabilities of EM to the evaluation of chemical composition of the specimen[4,5]. However, results have been limited and primarily focused on studying inorganic compounds within the cell structure, such as the presence of metals and other heavy ions[6–8]. The main challenge is that biological specimens possess considerable chemical and structural complexity while being mainly composed of a few light elements (carbon, hydrogen, oxygen, and nitrogen). As a result, EELS and EDS struggle to distinguish different biological molecules based on their compositional signatures.

Alternatively, one can use vibrational spectroscopy to distinguish between organic compounds, since the signature molecular vibrations correspond to the unique bonding configurations in compounds composed of the same elements[9]. Optical and beamline scattering techniques can provide this type of vibrational spectroscopy with ultrahigh spectral resolution[10,11], and when combined with scanning probes can attain nanoscale spatial resolution[12,13]. Such techniques are extremely powerful, but vulnerable to complex interactions between the probe tip and the sample, limiting the types of material that can be accessed with high spatial resolution[14]. EELS had been able to use the fine-structure of high energy core-loss states to perform some compositional analysis[4,15,16], but until recently, EM based spectroscopy lacked the energy-resolution to access these vibrational modes[17]. Modern monochromators in the scanning transmission electron microscope (STEM) have enabled single-digit-meV energy resolution in monochromated EELS (even down to ~3 meV/25 cm$^{-1}$ resolution under ideal conditions)[18]. More critically, this improvement in energy-resolution does not negatively influence the probe size, enabling simultaneously high spatial/spectral-resolution analysis with vibrational EELS[19–21]. As a result, new studies have used monochromated EELS to access traditionally inaccessible organic/biological materials such as amino acids[22,23], proteins [24], hydrogen-defects[25], and liquid water[26]. However, these studies have only been performed on isolated and purified samples (i.e. powders, flakes, liquid cells), and not on complex whole-cell biological specimens.

Here, we examine the vascular system of a cucumber stem using high-spatial-resolution vibrational EELS. The cucumber stem is a well understood material, and such plant-based specimens present an intriguing opportunity for well-defined cellular analysis at small length scales, due to the large variety in regions with unique characteristics. First, low-magnification defocused STEM imaging is used to identify the different cellular regions of the cucumber stem for direct analysis. Subsequently, vibrational monochromated EELS is used to selectively probe the cell walls and bodies within these different regions. Distinct vibrational signatures are identified, both for different regions within the vascular system, and between the cell walls and bodies within each region at the high spatial resolution of the STEM. To understand the vibrational signatures, we compare the whole-cell experiments to the vibrational spectra of the individual compounds that make up the cucumber stem (*i.e.* cellulose, hemicellulose, pectin, and lignin) as measured from monochromated EELS, optical Fourier-transform infrared (FTIR) spectroscopy, and first-principles density functional theory (DFT) calculations. We show that in specific areas of the vascular system, highly oriented cellulose chains experience enhanced coupling with the electron beam that allow us to unambiguously distinguish between the different cellular regions. These experiments clearly demonstrate that monochromated EELS opens a new pathway for observing the vibrational modes of individual cells, giving us the ability to probe light element bonding at a length scale of a few tens of nanometers.

The different types of cellular regions found in the cucumber stem are exemplified in Figure 1, which shows optical and electron microscopy images of the microtomed cucumber stem with the different cellular regions labelled. The optical microscopy image in Fig. 1a shows the entire sample and illustrates that the regions can be identified by the different cell body sizes/shapes and wall widths (i.e. the cambium and the phloem have similar cell sizes but different wall widths and different cell shapes).

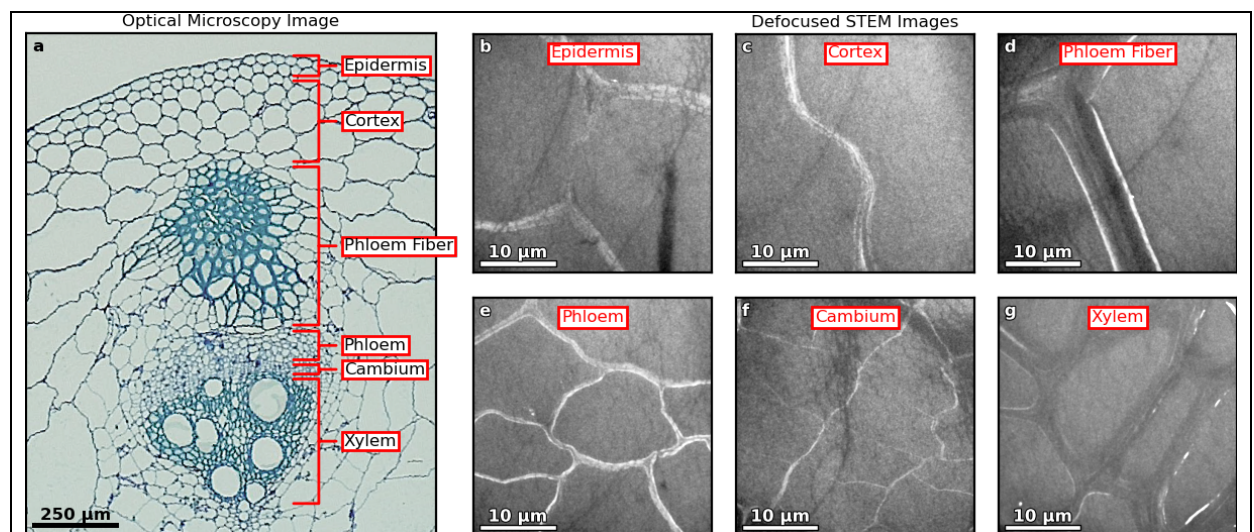

**Figure 1. Cellular Region Identification in Defocused STEM Imaging. (a)** Optical microscope image of a dyed cucumber vascular system with the different cellular regions labelled. Each region has its own unique combination of cell-size, cell-shape, and cell-wall thickness allowing identification from the much smaller FOVs accessible in STEM. **(b-g)** Defocused STEM images of the different cellular regions: (b) Epidermis, (c) Cortex, (d) Phloem Fiber, (e) Phloem, (f) Cambium, and (g) Xylem. From these images one can see the relative differences in the cell size and shape, allowing each area to be identified for STEM analysis.

The maximum STEM field of view (FOV) is far smaller than that of an optical microscope, and thus the whole cross section cannot be imaged simultaneously. However, using a low-magnification defocused bright field (BF) mode, individual ~30 μm segments can be imaged in the STEM, which is sufficient to determine which cellular region is being probed from the sizes of the cells, the width/character of the cell walls, and the relative positions of the regions with respect to one another. From the top to bottom there are six distinct regions: the epidermis, the cortex, the phloem fiber, the phloem, the cambium, and the xylem, as shown by the defocused STEM images in Fig. 1b-g. In the plant vascular system, the xylem, cambium, and phloem are of great interest as they provide nutrients and structural support to the plant.

A significant difficulty for the analysis of plant tissue samples with lateral dimensions on the order of millimeters with our electron probe (~1-2 Å in diameter) is the connection of the STEM FOV with the material superstructure. The process for localizing a STEM acquisition region with respect to the plant superstructure is outlined in Figure 2. Here, as an example of how the FOV is narrowed from ~30 microns to a suitable FOV for monochromated EELS, we show the full process for the acquisition in the cambium cell wall. We first use the defocused BF STEM image to identify the location of the cambium in the sample. The cambium has a unique cellular character, which is defined by thin-walled brick-like rectangular cells (Fig. 2a). After finding the cambium within the sample, we increase magnification in the defocused BF mode until the lacey carbon (~10 μm FOV) that supports the microtomed cross-section is visible, which allows us to take a reference map that includes both the lacey carbon and the cucumber stem cells (Fig. 2b). The random webbing of lacey carbon forms a unique pattern that can be identified even at much smaller FOVs, which is critical because monochromated EELS can only be performed in a *focused* STEM mode which has at best a 1-2 μm FOV. We then switch to a focused dark-field (DF) STEM imaging mode and correlate the patterns of the lacey carbon to the larger FOV defocused BF image to correlate the EELS-acquisition-region to its position within the overall cellular superstructure.

However, the primary challenge in plant tissue vibrational EELS is the inability to expose the delicate sample to the electron beam for long periods of time. The high energy electrons have orders of magnitude more energy than needed to break the chemical bonds of organics, causing the material to decompose under the electron beam. Furthermore, STEM sample preparation usually involves pre-treatments to mitigate hydrocarbon contamination (which can be deposited under the electron beam during an acquisition), such as plasma cleaning and high temperature annealing, which are not feasible for plant tissue due to the fragility of the material. Therefore, both beam damage and hydrocarbon contamination are present during experiments and preclude the conventional EELS practices used in robust inorganic systems with a single beam position and long dwell times.

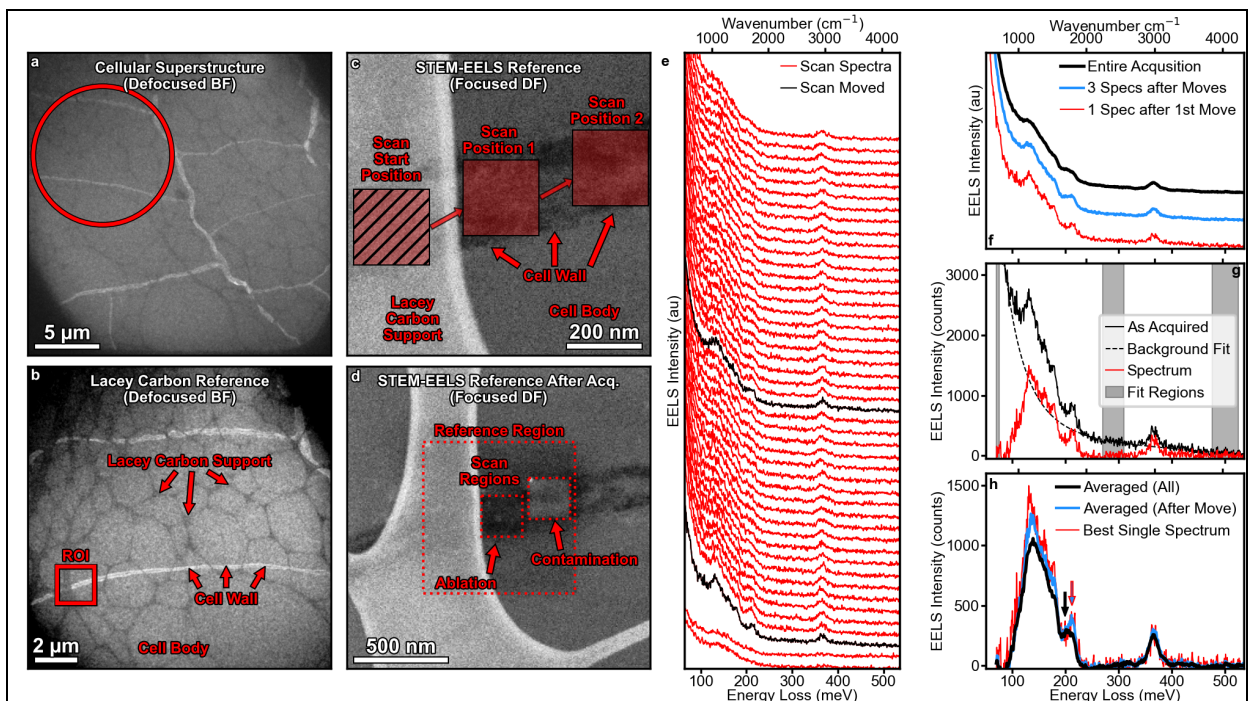

**Figure 2. Acquisition of Vibrational EELS in a Plant Tissue Sample.** **(a)** Large area defocused STEM image identifying the region as the cambium from its unique brick-like cells. **(b)** Defocused STEM reference with the lacey carbon support visible to help identify exact regions for EELS acquisition. **(c)** Dark field STEM image of area for EELS acquisition: 200 nm x 200 nm sub-scan box is selected and then a sequence of spectra is continually recorded while spreading the dose out over the sub-scan box. After the acquisition begins, the sub-scan box is moved to pristine region to acquire a spectrum from the specimen without any influence of beam damage. Such moves occur two or three times during the acquisition (here two moves). **(d)** STEM image of region after acquisition demonstrating electron beam damage: ablation – darker regions (indicating removed material), contamination – brighter regions (indicating deposited material). **(e)** The EEL spectrum recording with spectra acquired directly after moves highlighted in black (first recorded at the bottom, last at the top). **(f)** Average spectra for entire 50 spectrum recording, only the first three spectra after the two moves (so six spectra in total), and the single spectrum after the first move. **(g)** Demonstration of background subtraction to achieve pure vibrational signal. **(h)** Overlaid, averaged, background-subtracted spectra from (f), demonstrating that the influence of continued irradiation is a reduction in the sharpness of the vibrational peaks in most cases and in some even a change in the peak vibrational frequency (200-210 meV/1600-1700 cm$^{-1}$ region). Red/blue arrow denotes peak intensity in single and post-move spectra, black arrow denotes peak intensity across all spectra in recording.

Here, we use sub-scanning within a reference image to reduce the dose and the resulting contamination and damage. With a 2 Å diameter probe (spot size ~3 Å$^2$) and a beam current of 10 pA (beam conditions for these experiments), the areal dose for a 2 second spectrum acquisition would be ~4×10$^7$ e/Å$^2$. However, by sub-scanning over a 200 nm x 200 nm area during the acquisition, the effective areal dose per spectrum is significantly reduced to ~31 e/Å$^2$. The dose reduction provided by this strategy is optimal to obtain vibrational spectra that are not fundamentally altered by beam damage. This sub-scan approach also has the advantage that the size of the sub-scan box is an easily adjustable parameter, that can be reduced to increase the spatial resolution of the experiment or increased to further reduce the areal dose. Thus, one can choose the sub-scan to achieve the desired levels of signal-to-noise ratio (SNR) and spatial localization in the measurements.

This process is exemplified in Fig. 2c, which is now a focused DF STEM image as opposed to a defocused BF STEM image. To begin with, the sub-scan area is positioned on the lacey carbon (annotated with a hatched box labeled Scan Start Position), the idea being that last-second EELS optimization and tuning are performed outside of the target area. Then an EELS recording is initiated that consists of 40-50 sequentially acquired spectra with a dwell time of 2 seconds each. Once the recording has begun, the sub-scan is moved from the initial position onto the region of interest, which should still be pristine (Scan Position 1). However, even at the low effective dose rate offered by the sub-scan, damage does accrue over the course of several seconds. As a result, the sub-scanned region is moved a second time during the recording to a new pristine area (Scan Position 2), so there will be multiple spectra in each recording that contain measurements from pristine regions. The method allows for high spectral fidelity signatures to be obtained with high-energy EELS, and for the degradation mechanism in the sample to be understood by examining the acquisition chronologically. A key benefit of the strategy is exemplified in Fig. 2d, which shows a DF focused STEM image of the same region from Fig. 2c after the acquisition. Both beam damage (ablation) and hydrocarbon deposition (contamination) have resulted from different parts of the same acquisition. The presence of multiple failure mechanism necessitates careful data analysis to extract meaningful vibrational properties.

Each spectrum of EELS recording is shown in Fig. 2e chronologically from bottom to top. The first two spectra are acquired over the lacey carbon (hence the reduced signal intensity) and then the sub-scan is moved to the pristine region just before the $3^{rd}$ spectrum, resulting in significantly improved intensity and fidelity of the measured vibrational fine structure. By examining the spectra acquired afterwards, a gradual attenuation of the signal is observed as the damage/contamination mechanisms become more dominant. At the second move, another signal improvement is observed midway through the recording, accompanied by a second attenuation over the next several spectra. It is important to note that the signal is not as sharp as the one observed in the first move, indicating that beam damage or contamination has begun to accrue even outside of the sub-scan box during the first set of spectral acquisitions.

The influence on the measured vibrational response is depicted in Fig. 2f, where the average of the entire recording (top) is compared with the average of spectra acquired after the sub-scan moves (middle) and the best single spectrum acquired after only the first move (bottom). While averaging more spectra improves the SNR, it results in observable differences in the sharpness of the vibrational peaks. By subtracting the background (detailed in Fig. 2g) and directly overlaying the three spectra (Fig. 2h), it is observed that not only the peak sharpness is dramatically reduced but the peak frequency of some of the vibrational features also changes. In the 200 to 210 meV region (1600-1700 cm$^{-1}$), we see that the single spectrum is dominated by a single peak at 214 meV/1726 cm$^{-1}$ (marked with a black arrow), and in the averaged spectrum, the 214 meV/1726 cm$^{-1}$ peak is diminished and a new peak frequency is observed at 203 meV/1637 cm-1 (marked with a red and blue arrow). This shows that the damage/contamination mechanisms do not simply reduce the fidelity of the vibrational spectrum but fundamentally alter it.

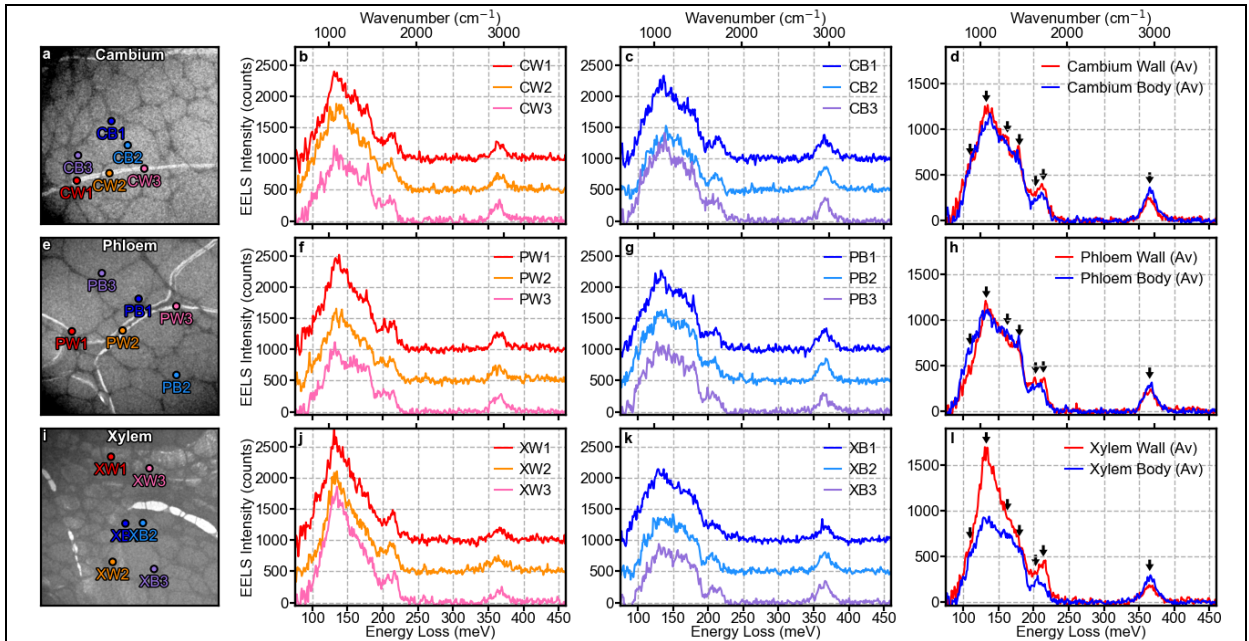

**Figure 3. Distinct Cellular Vibrational Signatures of the Cambium, Phloem, and Xylem. (a-d)** Analysis of the cambium region of the cucumber vascular system. (a) Defocused STEM image indicating regions from which the plotted spectra are acquired. (b) Comparison of three different EEL vibrational spectra acquired from the cambium cell wall. (c) Comparison of spectra acquired from cambium cell body. (d) Comparison of cell wall and cell body in the cambium with critical vibrational features highlighted with arrows. **(e-h)** Identical analysis shown in (a-d) for the phloem. **(i-l)** Identical analysis shown in (a-d) for the xylem. Subtle differences between the cell wall and cell body can be observed for each of the three regions, with the xylem demonstrating a clearly distinct vibrational response compared to the phloem and cambium. All spectra acquired with 200 nm x 200 nm sub-scans.

In Figure 3, we use the 'best single spectrum' (as defined in Fig. 2) to compare the vibrational response of the cell body and wall in three different regions of the sample: the cambium (Fig. 3a-d), the phloem (Fig. 3e-h), and the xylem (Fig. 3i-l). On the left (Fig. 3a, 3e, 3i) are the defocused reference images, where we have annotated three positions in the cell wall and three positions in the cell body. Here, we focus on the first spectrum at each position from the first sub-scan move to show the spectra least affected by beam damage, which are shown in Fig. 3b, 3f, 3j for the cell walls and Fig. 3c, 3g, 3k for the cell bodies. For each of set of three spectra, a common and unique vibrational fine structure is observed for the cellular region (despite some subtle spectrum-to-spectrum variation), which are averaged for the cell wall and cell body and compared in Fig. 3d, 3h and 3l.

All spectra exhibit a similar overall waveform, regardless of cell wall vs. cell body or different cellular regions within the cucumber vascular system: a dominant peak at 133 meV/1073 cm$^{-1}$, a secondary peak at 214 meV/1726 cm$^{-1}$, and an isolated peak at 365 meV/2944 cm$^{-1}$. The differences manifest mainly in the relative intensities of these peaks as well as the surrounding subpeaks and shoulders, which are observed at 110 meV/887 cm$^{-1}$, 163 meV/1315 cm$^{-1}$, 180 meV/1452/cm$^{-1}$, and 203 meV/1637 cm$^{-1}$. We have annotated the spectra in Fig 3d, 3h, and 3i with arrows to indicate the above frequencies.

By examining the EELS in all three different cellular regions (i.e. phloem, cambium, xylem), we can identify two classes of vibrational signatures: the features that can differentiate the cell walls from the cell bodies, and those that can differentiate the different cellular regions from one another. The signatures that differentiate the cell walls from the cell bodies are subtle but observed in all three cellular regions. They consist of a higher intensity of the 365 meV/2944 cm$^{-1}$ peak in the cell body, a higher intensity of the 203 meV/1637 cm-1 and 214 meV/1726 cm$^{-1}$ peaks in the cell wall, and a higher relative intensity of the 133 meV/1073 cm$^{-1}$ peak with respect to the surrounding 110 meV/887 cm$^{-1}$, 163 meV/1315 cm$^{-1}$, and 180 meV/1452/cm$^{-1}$ peaks in the cell wall.

However, by far the most significant signature is the cell wall vs. cell body comparison in the xylem with respect to the other regions. In the xylem, the 133 meV/1073 cm$^{-1}$ peak has double the intensity in the cell wall compared to the cell body, while in the phloem and the cambium, there is barely any enhancement, and it is only really relevant in terms of its comparison to the surrounding subpeaks and shoulders. More critically, the acquisitions are made with near identical beam-current, energy-resolution, sample-thickness, and dwell-time so there is no way for instrumental artifacts to generate such a huge intensity disparity, thus the enhancement must correspond to a real compositional aspect of the xylem that is not present in the phloem and cambium.

In order to understand the origin of these vibrational signatures, we seek to understand the vibrational modes that occur in the compounds that make up the cucumber vascular system, which are primarily cellulose, hemicellulose, pectin, and lignin[27,28]. Using powder samples of each compound we measure the vibrational spectrum using both FTIR optical spectroscopy and monochromated EELS, and the results are shown in Figure 4. We note in general that the selection rules for IR optical transitions and electron beam induced transitions are different, with the former typically limited to dipolar excitation and the latter capable of generating dipole excitations but also allowing for highly localized transitions which are forbidden in IR spectroscopy[29]. However, it is only systems which lack vibrational modes with strong dipole moments (such as crystalline silicon) where the forbidden transitions are dominant[30]. Thus, we expect our EELS measurements in the cucumber stem to be dominated by dipolar excitation and be largely comparable to FTIR.

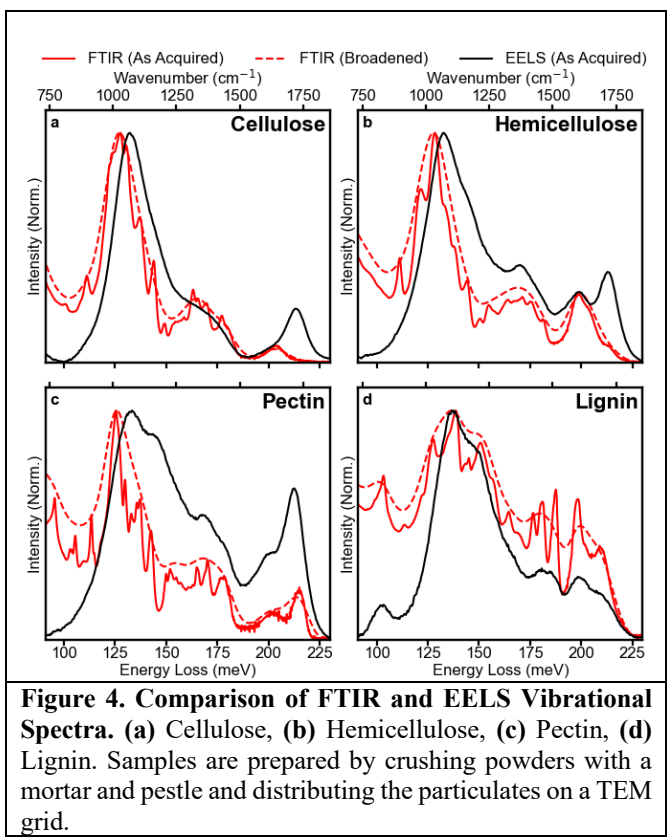

**Figure 4. Comparison of FTIR and EELS Vibrational Spectra. (a)** Cellulose, **(b)** Hemicellulose, **(c)** Pectin, **(d)** Lignin. Samples are prepared by crushing powders with a mortar and pestle and distributing the particulates on a TEM grid.

We focus here on the spectral regime between 90 meV and 230 meV (750 cm$^{-1}$ and 1800 cm$^{-1}$) as this is the regime with the biggest differences between cellular features as shown in Fig. 3. Here, the different compounds all show similar vibrational waveforms, both to each other and to the spectra acquired directly from the cucumber stem; with dominant peaks around the 133 meV/1073 cm$^{-1}$ EELS peak and various subpeaks and shoulders near the other EELS values. However, there are key differences between the FTIR and EELS that provide insight into the vibrational nature of the different peaks. The two most significant being the blue-shift observed in EELS with respect to FTIR of the dominant 133 meV/1073 cm$^{-1}$ peak in cellulose, hemicellulose, and pectin, *but not* in lignin, and the observation of a peak at 214 meV/1726 cm$^{-1}$ in EELS, but not in FTIR in both cellulose and hemicellulose.

**Table 1. Known Vibrational Modes for Cellular Compounds Associated with EELS Peaks.**
Key: $\nu$ – stretching, $\beta$ – in plane bending, $\delta$ – scissoring, $\omega$ – wagging, $\tau$ - twisting, $\gamma$ – torsional, as – asymmetric, s – symmetric, def – deformation, * – only observed in Raman, † – orientation-sensitive. All frequencies in cm$^{-1}$.

| EELS Peak | Mode Assignment | Lit. Frequency | Compound |
|---|---|---|---|
| 110 meV/887 cm$^{-1}$ | $\nu(COC)_{skeletal}$; $\tau^{\dagger}(CH_2)$[31] | 880-900; 971 | Cellulose |
| | $\nu(CO)_{ring}/\beta(COC)/\beta(CCO)/\beta(OCO); \nu_s(CO)_{ring}/\nu(CC)/\nu(CO)$[32] | 782; 841 | Hemicellulose |
| | $\nu(CC)$[32] | 889* | Hemicellulose |
| | $\nu(COC)_{skeletal}$[31] | 870-900 | Hemicellulose |
| | $\gamma(C-OH)_{ring}$; $\nu(COC)_{skeletal}$[31] | 816-817; 825-860 | Pectin |
| | $\gamma(C-OH)_{ring}$; $\delta(CCH)/\delta(COH)$[33] | 790, 830, 853; 888 | Pectin |
| 133 meV/1073 cm$^{-1}$ | $\beta(CCH)/\beta(CCO); \nu(CO)/\nu(CC)/\beta(COH)$[32] | 1001; 1121* | Cellulose |
| | $\nu_{as}^{\dagger}(COC); \nu_s(COC)$[31] | 1098; 1120* | Cellulose |
| | $\nu_{as}(COC)/\nu(CC)/\nu(CO)/\beta(CCH); \nu(CO)/\nu(CC)/\beta(COH)$[32] | 991; 1089 | Hemicellulose |
| | $\beta(CCH)/\beta(CCO); \nu(CO)/\nu(CC)/\beta(COH)$[32] | 999*; 1138* | Hemicellulose |
| | $\nu(CO)/\nu(CC)/\beta(COH)$[32] | 1086 | Hemicellulose |
| | $\delta(CCH)/\delta(COH); \gamma(COOH); \nu(CO)/\nu(CC); \nu(CO)+\delta(OH); \nu(COC)$[33] | 954; 990; 1034, 1050*,1119; 1085; 1156 | Pectin |
| 163 meV/1315 cm$^{-1}$ | $\beta^{\dagger}(CCH)/\beta^{\dagger}(CHO)/\beta^{\dagger}(COH)$[31] | 1378 | Cellulose |
| | $\tau(CH_2)$[32] | 1298 | Hemicellulose |
| | $\delta(CH)/\delta(OH)$[31] | 1256 | Hemicellulose |
| | $\delta(OH)_{COOH}; \delta(CH)$[33] | 1226; 1253, 1335 | Pectin |
| | $\omega(CH)/\delta(C-OH); (CH)_{def}/\beta(COH)$[31] | 1275; 1336 | Lignin |
| 180 meV/1452 cm$^{-1}$ | $\delta(CH_2)$[32] | 1473 | Hemicellulose |
| | $\delta(CH_2)$[32] | 1494 | Hemicellulose |
| | $\omega(CH_2); \delta(CH_2)$[32] | 1394; 1475 | Hemicellulose |
| | $\nu,\delta(C-OH)_{COOH}$[33] | 1403 | Pectin |
| 203 meV/1637 cm$^{-1}$ | $\nu(C=C)_{aromatic}$[31] | 1600-1610, 1659-1670 | Lignin |
| 214 meV/1726 cm$^{-1}$ | $\nu(C=O)$[31] | 1740-1755 | Hemicellulose |
| | $\nu(C=O)$[31] | 1740-1755 | Pectin |
| | $\delta(H_2O); \nu(C=O)_{COOH}$[33] | 1645; 1762 | Pectin |
| 365 meV/2944 cm$^{-1}$ | $\nu(CH); \nu_s(CH_2)$[32] | 2912; 2891, 2943 | Cellulose |
| | $\nu^{\dagger}(CH)$[31] | 2890-2900 | Cellulose |
| | $\nu(CH); \nu_{as}(CH_2)$[32] | 2930 | Hemicellulose |
| | $\nu(CH); \nu_s(CH_2)$[32] | 2938, 2955, 2967; 2997 | Hemicellulose |
| | $\nu(CH); \nu_{as}(CH_2)$[32] | 2916, 2970; 2938 | Hemicellulose |
| | $\nu(CH)$[31] | 2930 | Hemicellulose |
| | $\nu(CH)$[31] | 2945-2950 | Pectin |
| | $\nu(OH); \nu(CH)$[33] | 2653; 2942 | Pectin |

To understand the origin of these differences, and to understand how they correspond to the vibrational signatures observed in the whole cell sample, we review the literature to see which vibrational modes dominate the behavior of the different compounds. Table 1 shows a summary of several different literature sources on the vibrational modes in cellulose, hemicellulose, pectin, and lignin, and provides significant insights into the vibrational response of these materials[31–33]. The most critical aspect of Table 1 is the presence of orientationally sensitive modes in cellulose, which makes them highly sensitive to the polarization-selective dipole scattering of the electron beam.

Dipole scattering from an electron beam is predominantly polarized in the plane normal to the transmission direction, meaning the beam can only efficiently transfer momentum in directions radially outward from the beam position. As a result, highly polarized vibrational modes are enhanced when the beam is positioned such that it can transfer momentum directly into the polarization axis of the mode[34]. Especially in cellulose, which is formed of long straight chains of glucose monomers, the combination of the polarization sensitivity of the electron beam and the extreme anisotropy of the chain vs. non-chain directions in the cellulose should have a major effect on the measured spectrum.

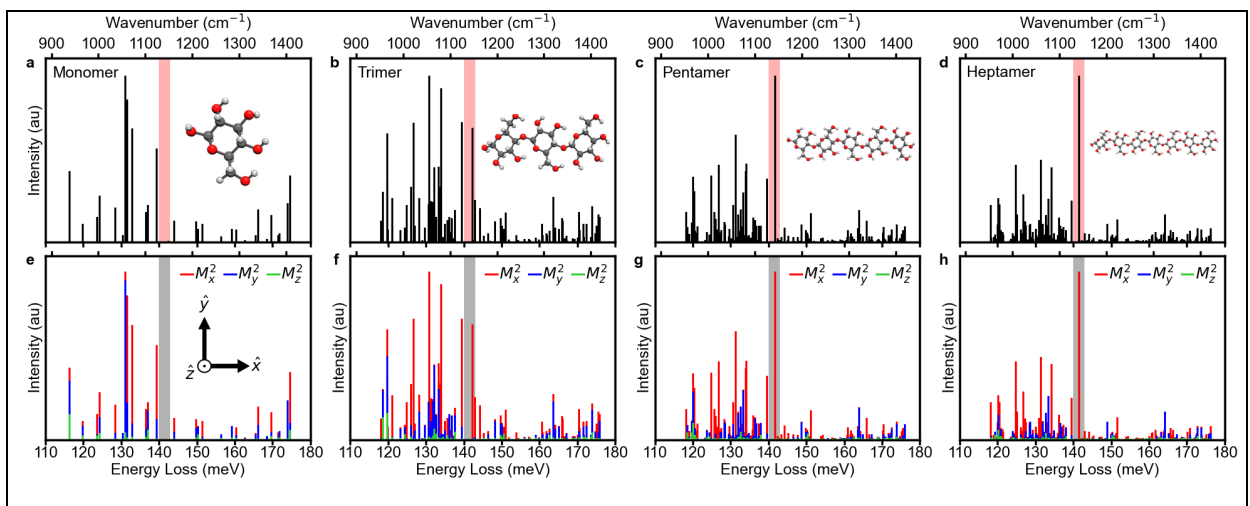

**Figure 5. Polarization Sensitive Modes in Cellulose. (a-d)** Vibrational eigenmodes of cellulose chains composed of one (a), three (b), five (c) and seven (d) glucose monomers, demonstrating an emergent chain mode (highlighted in red) that dominates the spectrum. **(e-h)** The square of *x*, *y*, and *z* components of transition dipole moments of each vibrational mode shown in (a-d) demonstrating that the dominant chain mode is the most polarization sensitive of all modes in the spectrum. The axes labels shown in the inset of (e) corresponds to the crystallographic directions of the monomer/oligomers shown in the insets of (a-d) with the chain-direction as *x*. Shaded areas (red in a-d, gray in e-h) indicate regions of emergent chain mode.

To determine the influence of orientation on the vibrational spectra, we perform first-principles DFT calculations of cellulose chains. By comparing oligomers of different lengths, we can tell what effect the rising anisotropy has on the vibrational response. Figure 5 shows the DFT calculations of eigenmodes for a glucose monomer (Fig. 5a), a trimer (Fig. 5b), a pentamer (Fig. 5c), and a heptamer (Fig. 5d). As the number of atoms in the calculations increases the number of modes increases, but the general distribution of modes remains the same in all four calculations. There is one significant exception: a mode at 141 meV/1137 cm$^{-1}$ begins to emerge in the trimer calculation (highlighted in red) which becomes more and more pronounced as the chain length increases. Moreover, this mode is quite close to the 133 meV/1073 cm$^{-1}$ frequency where we observed the enhanced intensity in the xylem wall.

Since the modes are determined computationally, the normal modes atomic displacement vectors are also known so we can directly visualize the vibration and assign it to a mode with DFT. Short videos of the highlighted chain vibration are shown in Extended Data Movies 1, 2, and 3.

From these we can see that these modes correspond to the asymmetric stretching of the glycosidic linkage of -COC-, denoted by $\nu_{as}$(COC), between the neighboring glucose monomers along the chain, the same mode predicted to be the dominant orientation sensitive mode for cellulose in the 133 meV/1073 cm$^{-1}$ peak regime in Table 1.

We note that the IR intensities for vibrational transitions between initial, $\Psi_{initial}$, and final, $\Psi_{final}$, vibrational states are proportional to the square of transition dipole moment, $M = \langle \Psi_{initial}(Q) | \vec{\mu}(Q) | \Psi_{final}(Q) \rangle$, carrying information about changes of a dipole moment, $\frac{\partial \vec{\mu}}{\partial Q}$, along given normal mode coordinate Q. We can decompose the vibrational spectrum into components corresponding to each cardinal direction of transition dipole moment, i.e. we can express the intensities shown in Fig. 5a-d as fractional sums of the transition dipole moments ($I = M_x^2 + M_y^2 + M_z^2$). The relative fractions of the intensity coming from the $x$, $y$, and $z$ dipoles are shown for the monomer (Fig. 5e), trimer (Fig. 5f), pentamer (Fig. 5g), and heptamer (Fig. 5h).

The transition dipole moments clearly show that the $\nu_{as}$(COC) has a large contribution in the $x$ (chain) direction. Many of the other modes present in the sample not associated with the $\nu_{as}$(COC) also have a much higher chain axis dipole moment than either of the out of plane directions. As a result, we can assign the enhanced intensity in the 133 meV/1073 cm$^{-1}$ peak in the xylem wall to the presence of highly oriented cellulose bundles in the xylem cell wall that are not present in the xylem cell body, or in the cell walls of the other regions of the cucumber vascular system. Furthermore, this conclusion is consistent with other studies that have reported on the xylem wall composition, which indicate that the xylem wall is mostly cellulose and that it is formed in bundles that run along the cell wall[31,35].

Such orientational effects can also explain the observed differences in the peak frequency of FTIR and EELS observed in Fig. 4 for cellulose, hemicellulose, and pectin, but not lignin. Since cellulose, hemicellulose, and pectin are polysaccharides, sugar monomers chained together by glycosidic bonds while lignin is not[27]. Other comparisons of FTIR and EELS have repeatedly seen noticeable shifts in peak frequency between the two techniques[22,23,25] and the preferential excitation of orientation sensitive modes in EELS (but not FTIR) is a highly likely origin.

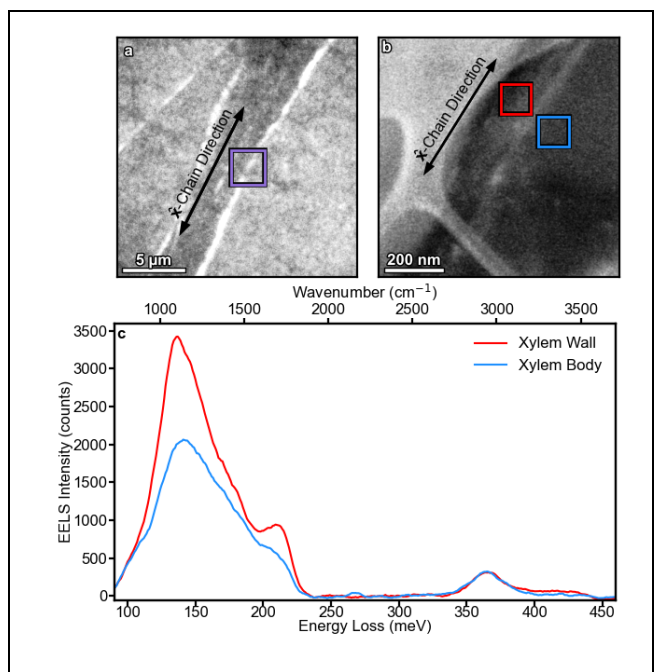

**Figure 6. Demonstration of nanoscale spatial resolution.** (a) Defocused STEM image highlighting area of xylem wall for analysis. (b) Dark field STEM reference image highlighting the interface between the xylem wall and body. Two adjacent 100 nm x 100 nm regions are scanned for EELS acquisitions in the wall (red) and body (blue). to the randomly oriented compounds in the cell body. (c) The respective spectra from the wall and body demonstrating the ability to clearly distinguish between different vibrational signatures at high spatial resolution.

Moreover, this orientation-sensitive vibrational signature can be identified with the superior resolution of the electron microscope. Figure 6 demonstrates the localization of the xylem wall vibrational signature. Fig. 6a shows a BF reference image of a xylem cell wall, and the Fig. 6b shows a focused DF STEM reference of the wall/body interface. Here, 100 nm x 100 nm scans on either side of the interface. The beam polarization direction extends radially outward from the probe position, meaning when the beam is inside the wall, we can efficiently excite vibration modes the chain ($x$) direction. As a result, the signatures from the chain modes are far more pronounced in the cell wall, compared to the cell body, as can be observed in the spectra shown in Fig. 6c, which originate from the 100 nm x 100 nm subscan regions denoted in Fig. 6b that are separated by only a few tens of nm.

The maximum spatial resolution here is limited more by the beam sensitivity of the sample as opposed to the actual localization of the signals. Here, the 100 nm x 100 nm subscan size is needed to reduce beam damage, and it could potentially be decreased in size even further to achieve even higher spatial resolution, but when the sub-scan is decreased the dose rate is increased making accurate high fidelity vibrational spectra more difficult to acquire. Thus, the careful tradeoff of needed areal dose vs. needed spatial resolution must be determined beforehand to ensure the optimal result.

Lastly, we can also highlight some of the challenges for vibrational EELS analysis of biological samples. Recall from Figure 4 that the 214 meV/1726 cm$^{-1}$ peak for pectin is strongly observed in both EELS and FTIR, while for cellulose and hemicellulose it is strongly observed EELS, but not (or just barely) observed in FTIR. From Table 1 we can see that the 214 meV/1726 cm$^{-1}$ definitively corresponds to $\nu$(C=O) stretching modes, which has a strong oscillating dipole moment and hence is strongly detected in techniques such as FTIR and EELS, but should only be present in pectin and hemicellulose, which is consistent with the FTIR measurements. However, the high intensity of the $\nu$(C=O) mode observed in the xylem wall is unexpected, since the composition is dominated by cellulose which should only possess C=O bond at the termination of their open-chain form, which makes up a small fraction of the material[31]. These inconsistencies arise because of the extreme sensitivity of the electron beam to localized heterogeneity, in this case surface oxidation of particulate samples used in the measurements.

The presence of the $\nu$(C=O) demonstrates that cellulose chains have been broken apart, which is known to occur when glucose oxidizes to form to gluconic acid or glucuronic acid[36]. The EELS measurements shown in Fig. 4 are conducted in the aloof mode where the beam is placed near to the surface of a particle and allowed to sample the vibrational modes evanescently, which makes the measurement very sensitive to surface effects on the face closest to the beam[37]. In the Extended Data Figure 2, we show a schematic of the aloof EELS acquisition, as well as the EELS/FTIR analysis of glucose, gluconic acid, and glucuronic acid, and show that the C=O stretch mode is observed in all three compounds for EELS, but only in gluconic and glucuronic acids for FTIR. Thus, we can see that EELS is sensitive to a thin surface layer that has oxidized on the powders, but FTIR is not because it mostly measures the bulk particle volume. Since surface oxidization has such a strong effect, it could also likely be the origin for the changes observed in the 200-210 meV (1600-1700 $cm^{-1}$) regime in Fig. 2, where we see that in the single spectrum the 214 meV/1726 $cm^{-1}$ peak is dominant but in the averaged spectrum it is dramatically reduced and the 203 meV/1637 cm-1 peak emerges, which is consistent with the ablation of an oxidized surface layer. Thus, the higher intensity of the 214 meV/1726 $cm^{-1}$ peak in the cell wall with respect to the cell body likely corresponds to an increased amount of cellulose in the cell walls with respect to the cell bodies (and hence a higher degree of oxidized gluconic acid).

In one regard, such a result is discouraging, as it makes deterministic assignment of vibrational modes to compositional origins challenging, but in other ways, it is encouraging. As it both demonstrates the validity of considering the individual pieces of a complex biological system to assess the composite whole, and the sensitivity to local disorder which is the primary purpose of using a high spatial-resolution technique such as monochromated EELS.

Conclusion

The energy-resolution available even in this latest generation of monochromated STEMs (3-5 meV/25-40 cm$^{-1}$) makes direct identification of the vibrational modes (which are often less than an meV in linewidth) a significant challenge. However, when combined with the spatial resolution of STEM, it holds great promise. Furthermore, as a transmission technique, monochromated EELS is not subject to the same surface sensitivity as scanning probe methodologies, making it ideally suited for topographically difficult or dirty samples. Moreover, there are excellent opportunities to improve spatial-resolution and spectral fidelity by combining with hybrid pixel detectors[38,39] (to reduce noise), advanced signal processing techniques to improve SNR while reducing dose rates[40–42], and vacuum-transfer techniques to reduce the influence of oxidization[43–45]. Vibrational EELS could also be combined with cryogenic STEM, to look at hydrated and even deuterated samples, where the nanoscale isotopic sensitivity could enable real space protein tracking to follow chemical reaction processes in other whole-cell samples. This unprecedented combination of spatial- and spectral-resolution opens many new pathways for plant tissue analysis and biology in the STEM.

Experimental:

*Sample Preparation*: Small pieces of the stem of a cucumber plant were fixed in 2% glutaldehyde (Sigma) in the phosphate-buffered saline (10 mM $Na_2HPO_4$, 1.8 mM $KH_2PO_4$, 2.7 mM KCl and 137 mM NaCl, Sigma) at pH 7.2. Subsequently they were rinsed with buffer and followed by water, and later dehydrated through a graded series of ethanol to 100%. Samples were gradually infiltrated with LR White resin to 100% final concentration and placed in gelatin capsules and polymerized in an oven at 50ºC. The samples were then sectioned via ultramicrotomy to create sections of the desired thickness (between 40-140 nm) and placed onto lacey carbon TEM grids.

The resin primarily cures around the edges of the stems and is not expected to penetrate into the center of the sample where the regions of interest (*i.e.* phloem, cambium, xylem) are situated.

*Electron Energy Loss Spectroscopy*: All EELS spectra were acquired on a Nion aberration-corrected high energy resolution monochromated EELS-STEM (HERMESTM) equipped with the Nion Iris Spectrometer operated at 30 kV, which potentially increases ionization damage[46] but also maximizes energy resolution. Experiments were conducted with a convergence angle of 27 mrad, collection angle of 25 mrad, and a beam current of ~10 pA. The energy resolution of the spectra is measured by taking full-width at half-maximum of the elastic scattering or zero-loss peak (ZLP) tail, which is between 9 meV and 15 meV for all spectra. In all acquisitions a third order exponential ($I(E) = A \cdot e^{aE^3 + bE^2 + cE + d}$) was used to fit and subtract the elastic tail of the vibrational spectrum, to better match the complex form of the highly monochromated ZLP tail.[47] Here we use three regions to fit the third order exponential: just before the onset of the dominant 130 meV peak (80 meV – 90 meV), in the region without any significant vibrational modes just after the 214 meV/1726 cm-1 peak (250 meV – 270 meV), and a final region after the upper limit of the O-H stretch mode (475 – 525 meV).

*Density Functional Theory Calculations*: DFT calculations were carried out with Q-Chem 5.4 package using B3LYP functional with 6-31G(d,p) atomic orbital basis set[48,49]. The structural models for the monomer through heptamer were constructed using the atomic coordinates from the experimentally determined crystalline structure of cellulose[50]. The minimization of the DFT ground electronic state was obtained with direct inversion of iterative subspace algorithm ($10^{-8}$ convergence threshold). The energy minimum geometries search was done with rational-function optimization in the internal coordinates. Following the structure optimization, the IR spectra were obtained from the analytical DFT hessians.

*Fourier-Transform Infrared Spectroscopy*: Small amount of powder (ca. 5 mg) was placed on the ATR crystal and the infrared spectrum was collected using PerkinElmer Frontier Fourier transform instrument.

**Acknowledgements**

Electron microscopy experiments were supported by the U.S. Department of Energy, Office of Science, as part of a user proposal at the Center for Nanophase Materials Sciences, which is a US Department of Energy Office of Science User Facility using instrumentation within ORNL's Materials Characterization Core provided by UT-Battelle, LLC, under Contract No. DE-AC05-00OR22725 with the DOE and sponsored by the Laboratory Directed Research and Development Program of ORNL, managed by UT-Battelle, LLC, for the U.S. Department of Energy. Theoretical calculations used computational resources of the Extreme Science and Engineering Discovery Environment (XSEDE) through allocation TG-DMR110037, and of the National Energy Research Scientific Computing Center (NERSC), a U.S. Department of Energy Office of Science User Facility located at Lawrence Berkeley National Laboratory, operated under Contract No. DE-AC02-05CH11231 using NERSC award BES-ERCAP0020403. BRE was supported by the U. S. Department of Energy Office of Science through the Genomic Science Program, Office of Biological and Environmental Research, under contract FWP ERKP752. FTIR experiments were supported by the U.S. Department of Energy, Office of Science, Office of Basic Energy Sciences, Separation Science program and Materials Chemistry program under Award Number DE-SC00ERKCG21 (IP and SJP).We would like to thank Professor Zheng-Hua Ye of the Plant Biology Department at the University of Georgia for providing the Cucumber stem sample.

# Supplemental Material for A New Path to Nanoscale Cellular Analysis with Monochromated Electron Energy-Loss Spectroscopy

*Jordan A. Hachtel[1*], Jacek Jakowski[2], Jingsong Huang[1], Santa Jansone-Popova[3], Ilja Popovs[3], Elizabeth A. Richardson[4], Barbara R. Evans[3], Peter Rez[5], Eric V. Formo[4*]*

[1] Center for Nanophase Materials Sciences, Oak Ridge National Laboratory, Oak Ridge, TN, USA
[2] Computational Sciences and Engineering Division, Oak Ridge National Laboratory, Oak Ridge, TN, USA
[3] Chemical Sciences Division, Oak Ridge National Laboratory, Oak Ridge, TN, USA
[4] Georgia Electron Microscopy, University of Georgia, Athens, GA, USA
[5] Department of Physics, Arizona State University, Tempe, AZ, USA
* Correspondence to: hachtelja@ornl.gov
* Correspondence to: eformo@uga.edu

**Extended Data Figure 1**

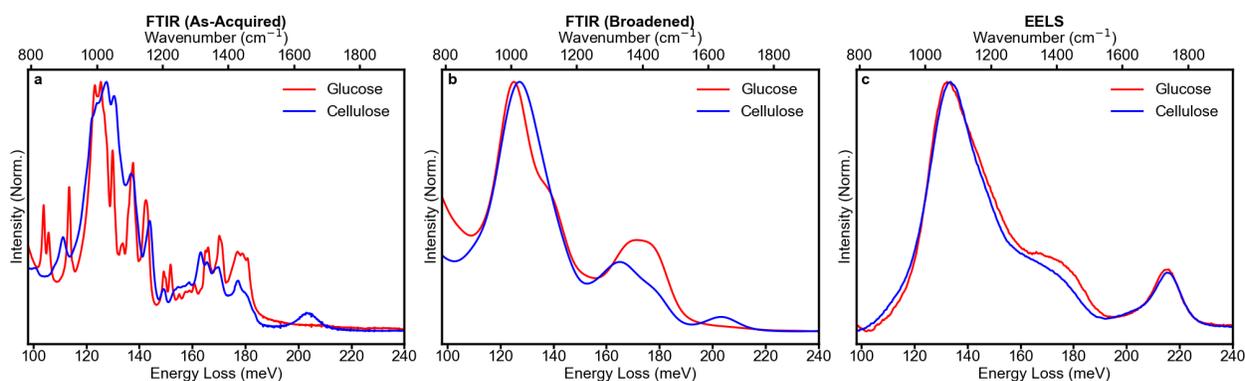

**Extended Data Figure 1. Comparison of Monomer Vibrational Spectrum to Compound Vibrational Spectrum.** To validate treating the larger compounds in the cucumber vascular system with the monomers that make up the compound we demonstrate a vibrational spectroscopy between cellulose and, its primary monomer, glucose. We use both Fourier-transform infrared spectroscopy (FTIR) and aloof monochromated electron energy-loss spectroscopy (EELS) to acquire vibrational spectra on powder samples. **(a)** Shows the FTIR comparison, **(b)** shows the FTIR broadened to match the EELS resolution ~6 meV, and **(c)** shows the EELS measurement. As we can see, while some small differences exist the general behavior and the dominant modes are very similar between cellulose and glucose, especially at the energy resolution accessible in an EELS experiment.

**Extended Data Figure 2**

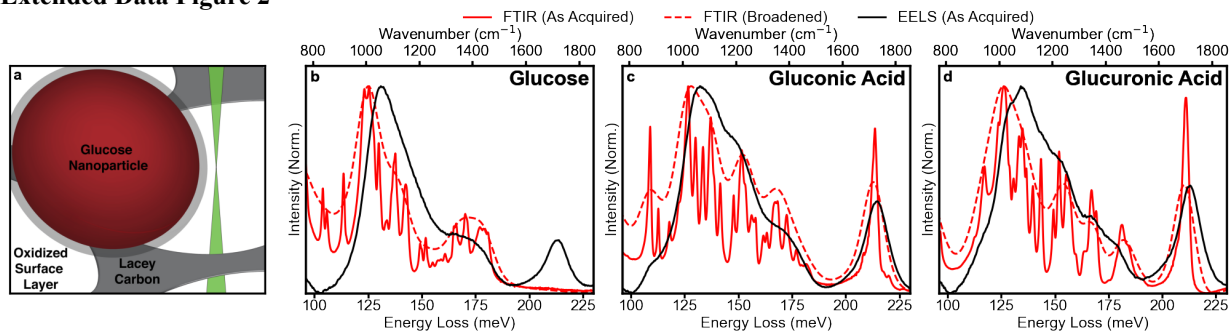

**Extended Data Figure 2. Oxidation in the Powder EELS Analysis. (a)** Schematic showing aloof EELS acquisition geometry. Glucose powdered samples are reduced to nanoparticles via mortar and pestle and dispersed on a lacey carbon grid. We find large particles hanging out into vacuum and acquire spectra through the aloof effect, by placing the beam 30-50 nm away from the sample, where the infrared vibrational energies can still be efficiently excited, but the damaging UV excitations cannot[1]. We perform aloof EELS experiments on glucose, gluconic acid, and glucuronic acid, and compare to FTIR samples on the as-acquired commercial samples analyzed with bulk FTIR. **(b)** EELS and FTIR of glucose showing (as observed for cellulose and hemicellulose in Figure 4 of the main text) a sharp peak at 214 meV in EELS with no corresponding peak in FTIR. This peak corresponds to oxidation of the glucose monomers in cellulose and hemicellulose which converts the glucose into an oxidized form such as gluconic acid or glucuronic acid. In these powders only a small surface layer is oxidized, which is why the peak does not show up in FTIR (as the oxidized surface makes up a negligible fraction of the total powder volume). However, in EELS the particulates are probed via the 'aloof' interaction, which preferentially samples the surface area causing the oxidized region to become dominant. **(c,d)** EELS and FTIR of gluconic (c) and glucuronic (d) acids both show the same sharp peak at 214 meV with the rest of the spectra appearing highly similar, confirming that this surface oxidization is likely the cause of the peak in Main Text Fig. 4. It is also of note that the blue shifting of the $\nu_{as}$(COC) peak is reproduced in both glucose and EELS, indicating that this is not an oxidation effect, but rather, a fundamental difference in the way EELS and FTIR excite orientational sensitive mode. In terms of the analysis of the more complex cucumber samples, the microtomed cross-sections mean that the EELS should primarily interact with the unoxidized bulk of the sample as opposed to the potentially oxidized surface effect, meaning that this effect is not likely to be a dominant effect on these spectra. However, by examining the three spectra in Fig. 2f of the main text (which correspond to different levels of beam damage) we can see the fine-structure of these peaks in the 200-210 meV spectral region changes as a function of beam-irradiation. Such changes could correspond to both additional oxidation, or the ablation of an oxidized surface layer from the sample, but in either case it is clear that unambiguous vibrational analysis in this spectral regime is a significant challenge.